\documentclass[sigconf]{acmart} 

\usepackage{tikz}
\usepackage{enumitem}
\usepackage{svg}
\usepackage{fancyvrb}

\usepackage{longtable}

\newcommand{\code}[1]{\tikz[baseline=(X.base)]\node [draw=black,fill=gray!10,semithick,rectangle,inner sep=2pt, rounded corners=3pt] (X) {\small\textsf{#1}};}

\newcommand{\codeH}[1]{\tikz[baseline=(X.base)]\node [draw=black,fill=gray!10,semithick,rectangle,inner sep=1pt, rounded corners=3pt,minimum height=14pt] (X) {\small\textsf{#1}};}

\newcommand{\nrc}[1]{#1}

\newcommand{\new}[1]{#1}

\newcommand{\tool}[1]{\textsc{#1}}
\newcommand{\ignore}[1]{}

\title{A Systematic Mapping Study of Code Quality in Education -- with Complete Bibliography}

\author{Hieke Keuning}
\affiliation{
  \institution{Utrecht University}
  \country{The Netherlands}
}
\email{h.w.keuning@uu.nl}

\author{Johan Jeuring}
\affiliation{
  \institution{Utrecht University}
  \country{The Netherlands}
}
\email{j.t.jeuring@uu.nl}

\author{Bastiaan Heeren}
\affiliation{
    \institution{Open University of the Netherlands}
    \country{The Netherlands}
}
\email{bastiaan.heeren@ou.nl}

\setcopyright{none}
\renewcommand\footnotetextcopyrightpermission[1]{} 


\begin{document}

\begin{abstract}
While functionality and correctness of code has traditionally been the main focus
of computing educators, quality aspects of code are getting increasingly
more attention. High-quality code contributes to the maintainability of software systems,
and should therefore be a central aspect of computing education.
We have conducted a systematic mapping study to give a broad overview of the research 
conducted in the field of code quality in an educational context. The study
investigates paper characteristics, topics, research methods, and the targeted
programming languages. We found \nrc{195} publications (1976--2022) on the topic in multiple databases,
which we systematically coded to answer the research questions.
This paper reports on the results and identifies developments, trends, and new opportunities
for research in the field of code quality in computing education. 
\end{abstract}


\maketitle
\pagestyle{plain} 

\section{Introduction}
Software quality is an important subject that Computer Science students need to
learn during their studies.
The quality of code, considering aspects
such as naming, documentation, layout, control flow and structure,
contributes to the readability, comprehensibility, and maintainability of software.
Code style and quality is often discussed in the context of other software engineering
topics, such as testing, code reviewing, and quality assurance (QA) in general. Automated assessment tools
and tutoring systems might give feedback on code style, besides the correctness of
solutions. Code quality
has historically
not been the main focus of educators~\cite{1201,927}, possibly due to time,
workload, lack of knowledge, and perceived lower importance.
However, we have noticed an increase in interest in this topic, which we may (or may not)
confirm with this study.
 

The goal of this Systematic Mapping Study~\cite{petersen2008systematic} is to identify the landscape of studies that have
been conducted on code quality in education. To our knowledge, this is the first
overarching study on this topic. We first identify publication characteristics such as year,
venue (journal/conference), followed by the topics, methods, languages, and relevant technical aspects.
\new{This paper makes the following contributions: 
(1) a large and complete list of papers on the topic,
(2) a broad overview of the research area, 
and (3) an identification of research trends and new research opportunities.}

\textit{This paper is an extension of a conference paper~\cite{sms-iticse}, and includes the complete reference list and coding.}

\section{Background}
\label{sec:background}

In this section we give our definition of code quality for this study, and describe several
relevant terms and aspects. We also briefly discuss other mapping
studies related to computing education.

\subsection{Terms and definitions}
\label{sec:terms}

\textit{Software quality} and \textit{code quality} are sometimes intertwined,
however, we consider code quality to be a more specific aspect of software quality.
The ISO/IEC 25010 standard for software product quality comprises eight quality characteristics,
among which functional suitability, usability, reliability, and maintainability.
The last characteristic can be subdivided into modularity, reusability, analysability,
modifiability, and testability.
High-quality code can contribute to these characteristics.


Code quality is a term without a clear meaning and with various interpretations.
We choose to focus on code quality as an aspect that appears \textit{after} writing
the initial program, dealing with analysing, reflecting on, and improving the program's static
characteristics.
We are interested in properties of source code that can be observed directly.
As such, we focus on the \textit{static} properties of code, as opposed to the
\textit{dynamic} properties such as correctness, test coverage, and runtime performance.
We focus on the categories from the rubric designed by Stegeman et al.~\cite{581}
to assess the quality of student code.
These categories are documentation, layout, naming, flow, expressions, idiom, decomposition,
and modularization. 

Problems with these aspects are often denoted as \textit{code smells}, a term introduced
by Fowler~\cite{Fowler1999}. Code smells may
indicate a problem with the design of functionally correct code, affecting quality
attributes of the software. Examples are duplication, dead code, overly complex code,
and code with low cohesion and high coupling.

\textit{Refactoring} is improving code step by step, while its functionality stays the same.
Fowler's~\cite{Fowler1999} well-known book describes a collection of refactorings, such as
extracting a class or method, introducing an explaining variable, pull up a field or method,
and replacing a magic number with a symbolic constant.
\textit{Design patterns} are reusable solutions to common problems in code~\cite{gof},
and can be used when refactoring.



To support developers with analysing and improving their code, many tools and systems
are available. 
Tools such as PMD, Checkstyle, SonarQube, Resharper, and linters can
automatically detect and report quality issues and code smells in a program.
These tools often employ static analysis techniques to analyse code, although static
analysis has a broader application and can also be
used to identify bugs and errors.
There are also many tools to support the refactoring of code, often integrated in modern
IDEs.


\subsection{Related work}

\textit{Systematic literature reviews}, which dive deep into the literature on some topic,
are increasingly being conducted for Computing Education (CEd)
topics (see section~\ref{sec:relfields} for examples related to our topic).
A \textit{systematic mapping study} aims to give a broad overview of a particular research
area, usually by categorising its publications. While mapping studies are common
in medicine, they are less common in other fields, such as software engineering~\cite{petersen2008systematic} and CEd. 
A systematic mapping study on software testing in introductory programming courses
by Scatalon et al.~\cite{scatalon2019software} is the most relevant to our study,
because they also investigate a software quality aspect in an educational
context. The
authors selected 293 papers and categorised them on their
topic and evaluation method.

Numerous mapping studies and literature reviews have been conducted on topics
related to code quality, such as a mapping study on source code metrics~\cite{nunez2017source}, and
a tertiary review on smells and refactoring~\cite{lacerda2020code}. However, these studies are not aimed at
code quality in the context of \textit{education}, the topic of our study.

\section{Method}
\label{sec:method}


We generally follow the process from Petersen et al.~\cite{petersen2008systematic} for doing systematic
mapping studies in software engineering. We employ a different
approach to classifying studies, as described in section~\ref{sec:coding}.

\subsection{Scope and research questions} 
\label{sec:scope}

The scope of this mapping study is:
\begin{quote}\textit{
Research on educational activities and support concerning code quality (as defined
in~\ref{sec:terms}),
such as: instruction, analysis, assessment, tool support, tasks, and
feedback.}
\end{quote}

\noindent Within this scope we address the following research questions:
\begin{description}
    \item[\textbf{RQ1}] Where are the papers published?
    \item[\textbf{RQ2}] Which topics have been addressed?
    \item[\textbf{RQ3}] Which types of studies have been conducted?
    \item[\textbf{RQ4}] For which programming language is the intervention? 
    \item[\textbf{RQ5}] What are the trends over time?
    \item[\textbf{RQ6}] Which other topics are closely related to code quality?
\end{description}

\subsection{Search process} 

The inclusion and exclusion criteria are defined in table~\ref{table:criteria}.
We have first assembled a base list of 40 papers that have been collected by ourselves
over the years and meet the criteria. Two authors have verified that all publications
on the list should be included.

\subsubsection{Database search}
We collected the keywords from the base papers, removing very general terms such
as `university' and `examples', very specific terms such as names of tools and programming
languages, and terms indicating the type of study. Based on these keywords we experimented
with various search strings, checking whether the papers would end up in the search results.
Because code quality is defined and named in various ways, we have used several
specific terms in the search string to be as inclusive as possible. During the process we have made the scope more clear by explicitly defining
the edge topics for RQ6, as discussed in section~\ref{sec:relfields}.

We chose three databases, Scopus, ACM and IEEE, which cover a wide range of publications
and allow searching with a complex search string.
\new{The search includes papers up to and including 2022. Because the final searches were conducted in
December 2022, a few papers from 2022 could be missing.}
The final search string is shown below. We applied the search string to title, abstract,
and keywords, and made some adjustments to match
the database requirements. From our base list of 40 papers, 36 were found by this search.


\begin{Verbatim}[frame=single,rulecolor=\color{gray},fontsize=\small,framesep=0.5mm]
    (programm* OR code OR coding OR software) 
AND ("code quality" OR "software quality"
     OR "design quality" OR refactoring 
     OR "static analysis" OR "software metrics"
     OR smell OR readability
     OR "code style" OR "coding style"
     OR "programming style") 
AND (student OR teach* OR educat* OR curriculum OR novice)
\end{Verbatim}

\begin{figure*}[bt]
	\begin{center}
	\includegraphics[width=16cm]{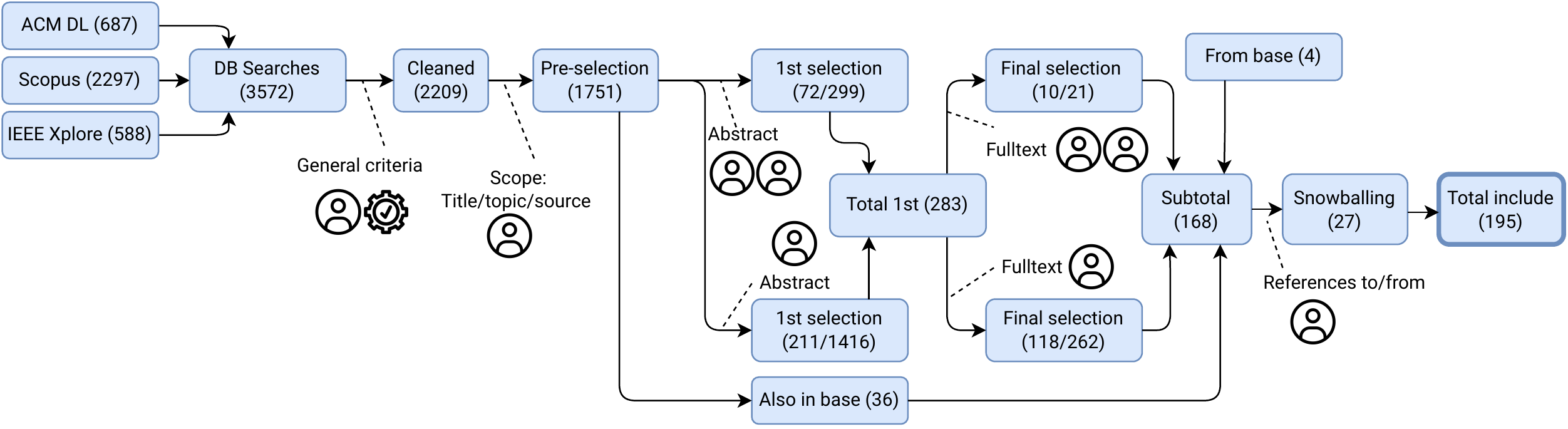}
	\end{center}
	\caption{Paper selection process; person icon denotes one/two author(s); cog icon denotes
	an automated step. } 
	\label{img:flow}
\end{figure*}

\begin{table*}[t]
\caption{Inclusion and exclusion criteria.}
\label{table:criteria}
\renewcommand{\arraystretch}{1.4}
\footnotesize\sffamily
\begin{tabular}{l p{7cm} p{8cm}}
    & \textbf{Include} & \textbf{Exclude} \\\hline
General 
    & Scientific publications (journal papers and conference papers) in English. 
    & Posters, papers shorter than four pages, theses, technical reports, books.
      Papers we cannot find. Papers preceding an extended version of a paper. \\

Topic 
    & Publications that describe interventions in a formal educational context
(high school/K-12, higher education). 
   & Educational contexts aimed at professionals working in practice.
Publications on automated feedback tools that provide style feedback
alongside other error feedback, with no particular focus on code quality.
Studies in which students are (among) the participants, but with no particular focus
on education.  
Plagiarism. Static analysis tools used for assessing correctness etc.            \\

Language 
    & Code and design of general-purpose programming languages, teaching languages (e.g.~Scratch).               
    & Domain-specific languages such as SQL, low-level programming, very specific contexts (e.g. shader programming, or block-based robot programming).  \\           

Focus 
    & A substantial part of the paper should be on code quality.                
    & Interventions that only lead to improved code quality (among others), but are not
specifically about code quality (further discussed in \ref{sec:relfields}).
\\

Quality
	& Publications should be indexed in Scopus, the ACM library, or IEEE library.
	Exceptions can be made for highly cited papers. 
	& 
 \\
 \hline 
             
\end{tabular}
\end{table*}

\noindent Next, we elaborate on the process steps
 (summarised in figure~\ref{img:flow}).

\vspace{-7pt}

\paragraph{Cleaning} One author combined the results from the three data-bases, and removed entries
that are not papers, or are too short, and deleted duplicates based
on title automatically using a script.

\vspace{-7pt}

\paragraph{Pre-selection} One author filtered the list for exclusion based on title,
and/or publication source, which were obviously out of scope because they are not about code quality and/or
educational setting.

\vspace{-7pt}

\paragraph{Selection} Two authors assessed a subset of the remaining
list by reading title/abstract/keywords and selecting yes/no/maybe.
Both `yes' and `maybe'
indicated that we will consult the fulltext. If only one of the authors selected
`no', we discussed whether or not the fulltext should be consulted. We had three rounds of around
100 papers each, with an agreement of 77\%, 78\%, and 89\%, respectively.
One author assessed the remaining papers. For the fulltext selection we also checked and discussed
several papers with two authors, after which one author selected the remaining papers.
After this step, we had a selection of \nrc{168} papers for inclusion in our study.

\subsubsection{Snowballing} The ambiguity surrounding the definition of code quality
prevents constructing a search string that finds all relevant research. To find
additional publications, we have performed \textit{snowballing}: identifying
relevant references from (backwards) and to (forwards) a set of papers~\cite{wohlin2014guidelines}.
For all \nrc{168} papers found in the previous steps, one author inspected all
references from and to the paper (the latter using Google Scholar), and selected those within
the scope. This inspection of thousands of references led to \nrc{27} 
additional papers. We stopped after one round of snowballing; we believe a second
round would unlikely yield more relevant papers, because these papers would not have cited
any of the papers from the database search. To answer RQ6, we kept a record of topics
and papers referred to during snowballing that were outside our scope.
  

\ignore{There is no obvious method to select a starting set of papers.
As the start set we choose papers from our base list, which we sorted
by Google Scholar citation count. Highly cited papers will lead to many relevant citing
papers, and we expect that popular papers will have a good overview of related work.
We then selected ten papers while ensuring the set adheres to Wohlin's
characteristics of a good start set: diversity in venue, year, authors, and community.
Our starting set was ~\cite{todo}. }

\subsection{Coding} 
\label{sec:coding}

To answer RQs 2--4, we coded each paper in four categories: topic, aspect, method, and language.
Codes are shown in a \code{box}.
We use the topics from the mapping study on testing
in programming courses by Scatalon et al.~\cite{scatalon2019software} as our base
for RQ2 (topic), with some small adjustments to fit our scope. 
We assigned one topic to each paper, representing its main focus or goal.

\begin{itemize}[leftmargin=*]
 \item \code{Curriculum} The integration of code quality in the computing curriculum as a whole or in individual programming courses.
\item Instruction:
    \begin{itemize}
        \item \code{Course materials} 
        \item \code{Programming assignments} In addition, guidelines to conduct assignments related to code quality. 
        \item \code{Programming process}
        \item Digital tools, either an \code{external tool} or \code{selfmade tool}.
        \item \code{Teaching methods} Used when a paper addresses multiple of the instruction elements above.
    \end{itemize}
\item Learning outcome:
    \begin{itemize}
        \item \code{Program quality} Assessment of students' submitted code.
        \item \code{Perceptions} Students' (or teachers') attitudes towards code quality.
        \item \code{Behaviours} Programming/refactoring behaviour. 
        Broader than just program quality, but may include it.
        \item \code{Concept understanding} Assessment of students' knowledge of code quality concepts.
     \end{itemize}
\end{itemize}

\noindent For RQ2 we also identified for each paper whether it deals with one or more of the
following domain-specific subtopics, attached when a topic is
present in the title, abstract, or keywords of the paper:
\code{design patterns}, \code{refactoring},
\code{code/design smells}, 
\code{static analysis},
\code{readability}, and
\code{metrics}. 
These terms were taken from the keyword analysis, and the term `readability' was
most often used to refer to code quality by developers, educators, and students~\cite{927}.

For the method (RQ3) we used categories from a recent Computing Education conference (Koli Calling 2021),
shown below. 
A paper will only be coded by its main method.

\noindent 
\setlength{\tabcolsep}{1pt} 
\begin{tabular}{l l l}
\codeH{Literature review} & \codeH{Qualitative} &
\codeH{Case study}  \\ [3pt]
\codeH{Descriptive/correlational} & \codeH{Survey} &
\codeH{Quantative/other}  \\ [3pt]
\codeH{System/Tool report} & \codeH{Theory paper} &
\codeH{Experience report}  \\ [3pt]
\codeH{(Quasi-)experimental} & \codeH{Mixed methods} & \codeH{Discussion paper}\\ [2pt]
\end{tabular}

A subset of 11 papers were individually coded in all four categories 
by two authors, after which differences
were discussed and resolved. One author coded the remaining papers.


\section{Results and discussion}
\label{sec:results}

The full list and coding of the \nrc{195} papers can be found online in a searchable
table\footnote{\url{www.hkeuning.nl/code-quality-mapping}} and in table~\ref{table:full}.
This section aggregates the findings and highlights examples from each category.

\subsection{Paper characteristics (RQ1)}
\begin{figure}[bt]

	\begin{center}
	\scriptsize\sffamily
	\includesvg[width=0.9\columnwidth]{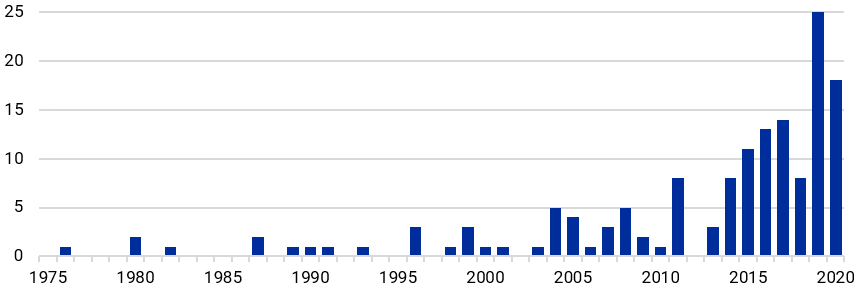}
	\end{center}
	\caption{Number of publications per year.}
	\label{img:pubs}
\end{figure}

Figure~\ref{img:pubs} shows the publication years of the papers.
The first publication appeared in 1976, but publications were
rare in the 70s and 80s. After increasing only slightly in the 90s and 2000s,
the attention for code quality in education clearly has been rising in the last decade.

Table~\ref{table:pub} shows the main journals (\nrc{33} papers) and conference
proceedings (\nrc{161} papers) in which the papers were published. The most common venues
are, as expected, related to Computing Education, however, the diversity of
the remaining venues is broad. Papers on the topic have been published in venues
on human-centric computing, software engineering, games, educational technology,
program comprehension, and others.


\begin{table}[!tb]
\caption{Publication venues (conference/journal) with at least 3 publications.}
\label{table:pub}
\footnotesize\sffamily
\begin{tabular}{l l l r}
\hline
\textbf{Name}  & \textbf{C/J}& \textbf{Type} & \textbf{\#} \\
\hline
Special Interest Group on CS Ed. (SIGCSE) & C & Computing Education & 21\\
Innovation and Technology in CS Ed. (ITiCSE) & C & Computing Education & 11  \\
Frontiers in Education (FIE) & C & Engineering Education & 11 \\
Koli Calling & C & Computing Education & 7 \\
Australasian Computing Education (ACE) & C & Computing Education & 7 \\ 
Computer-supported education (CSEDU) & C & Educational Technology & 5 \\ %
Computing Sciences in Colleges & J & Computing Education & 4\\
Visual Languages and Human-Centric Computing & C & Human-Centric Comp. & 4\\ 

IEEE Blocks \& Beyond & C & Block programming & 3 \\
Systems and Software & J & Software Engineering & 3 \\
IEEE Access & J & General computing  & 3 \\
Learning @ Scale & C & Educational Technology & 3 \\
 \hline 
             
\end{tabular}
\end{table}

\subsection{Topics (RQ2) and methods (RQ3)}

\begin{figure}[bt]
	\begin{center}
	\includegraphics[width=1\columnwidth]{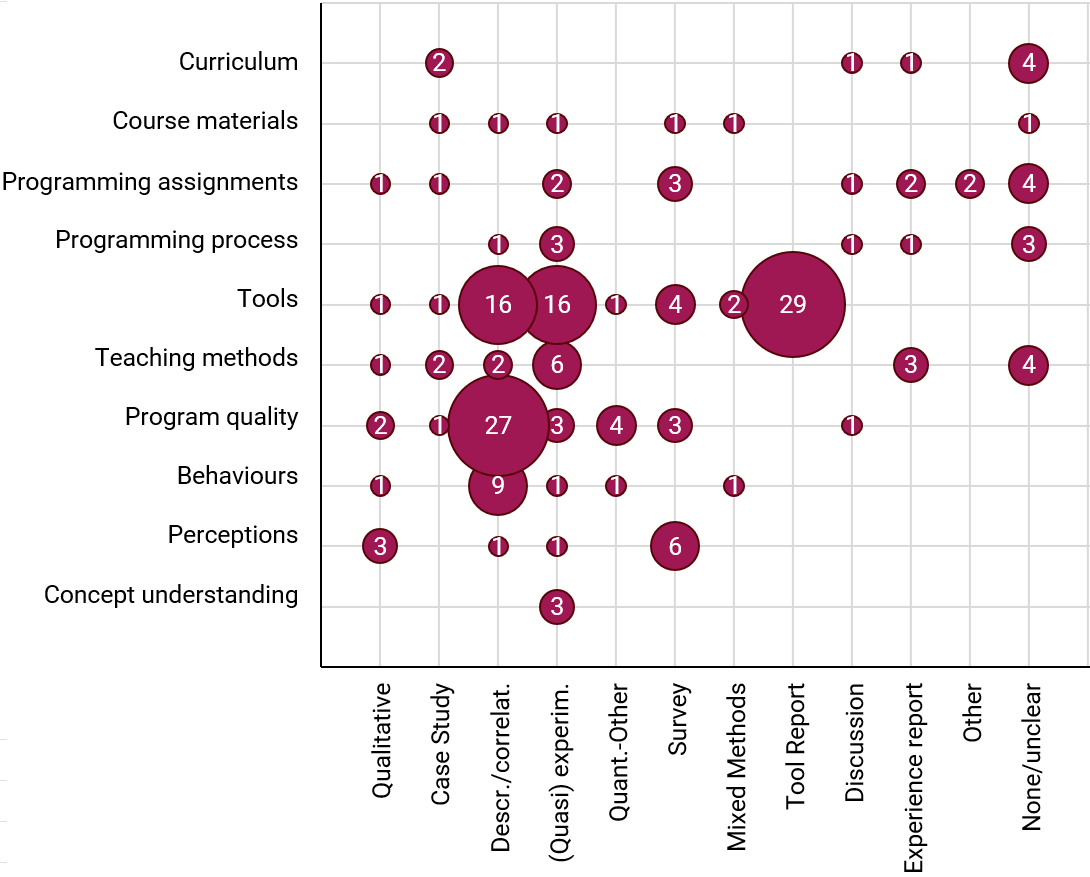}
	\end{center}
	\caption{Correlation matrix of topics and methods.}
	\label{img:topic-methods}
\end{figure}

\begin{figure}[bt]
	\begin{center}
	\includegraphics[width=0.8\columnwidth]{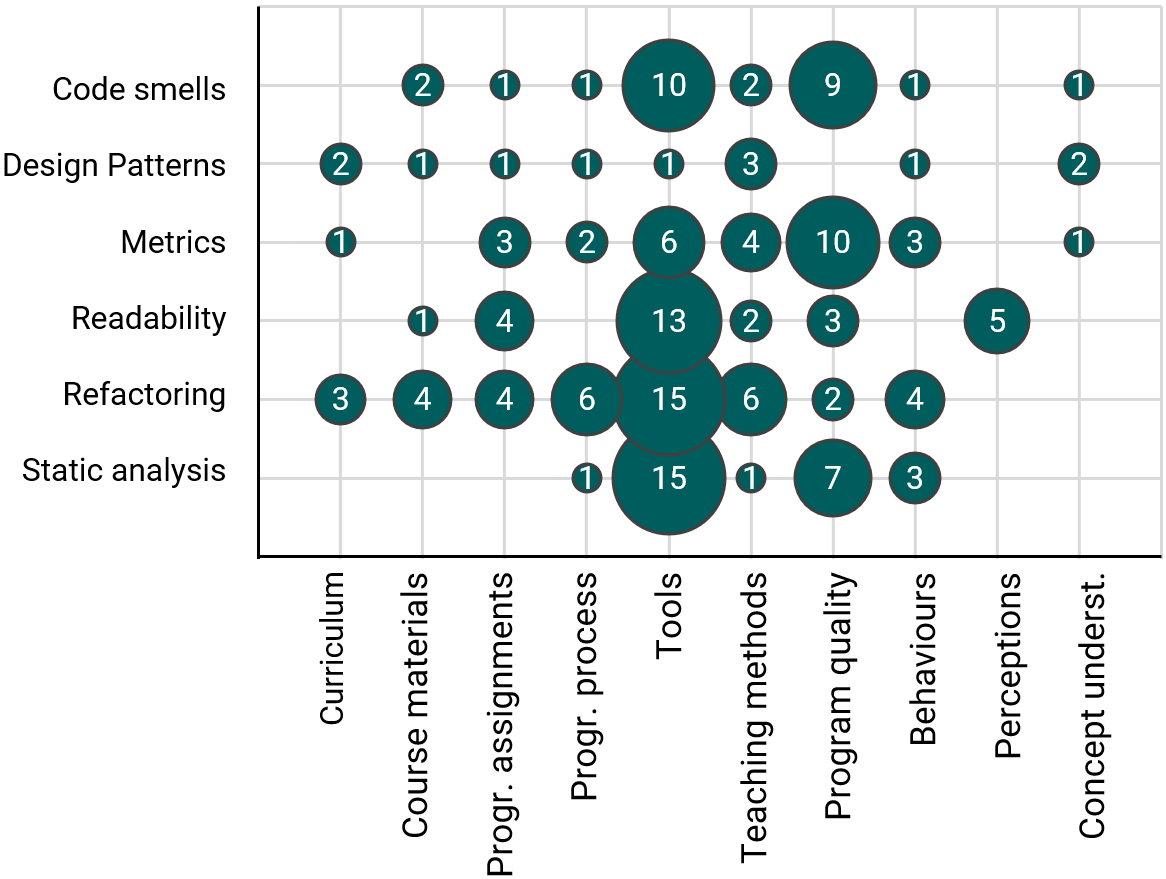}
	\end{center}
	\caption{Correlation matrix of topics and code quality aspects (a paper can have more than one aspect).}
	\label{img:topic-aspects}
\end{figure}

Figure~\ref{img:topic-methods} shows the correlation matrix of the topics
and the methods. We did not find any literature reviews or theory papers, therefore these categories
were omitted.
We notice two major topics: program quality (\nrc{41} papers)
and tools (\nrc{70} papers). We have made a distinction between tools created by
the authors (\nrc{59}), and the use of an external tool (\nrc{11}).
Figure~\ref{img:topic-aspects} shows the correlation matrix of the topics
and code quality aspects. All aspects clearly appear in multiple papers.

\subsubsection{Curriculum}

We have found only \nrc{eight} papers that revolve around integrating code quality
into the \code{curriculum}. As an example, Kirk et al.~\cite{1201}
study the prevalence of code quality in introductory programming courses.
Techapalokul and Tilevich~\cite{712}
advocate for the importance of integrating code quality in the teaching of
programming in block-based environments,
even though this code is usually not intended for practical use.
Haendler and Neumann~\cite{47} present a framework for the assessment and training of
refactoring.

\subsubsection{Instruction}

Overall, we observe that code quality in education revolves for a large
part around digital \code{tools}. Looking at the code quality aspects, we notice that
those tools focus on several of them, such as identifying code smells and
refactoring code, often using static analysis techniques. 
\tool{AutoStyle}~\cite{1479} gives data-driven feedback on how to
improve the style of correct programs step by step. Other recent tools are
\tool{CompareCFG}~\cite{487}, and a Java critiquer for antipatterns~\cite{366}.
\tool{RefacTutor}~\cite{1432} is a tutoring system to learn refactoring in Java.
\new{Keuning et al.~\cite{n297} present a tutoring system in which students practice
with improving functionally correct code, with the help of automated feedback and hints. }

In some tools a `gamification' approach was taken. Zsigmond et al.~\cite{853}
present a system in which badges are awarded to students who adhere to good coding
standards, using \tool{SonarQube} for static analysis. Examples of badges are 
`doc ace', `complexity guru', and `stylish coder'. 
\tool{Pirate Plunder}~\cite{1951} is a game in which children learn to
investigate and fix code smells in a Scratch environment.

We have also found papers on tools that focus on very specific aspects of
code quality. For example, the \tool{Earthworm} tool gives automated suggestions
of decomposition of Python code~\cite{647}. \tool{Foobaz} gives feedback on variable
names~\cite{824}. 
\new{Charitsis et al.~\cite{n337} present a system based on machine learning techniques
that can detect poor function names and suggest improvements.}


Examples of external (professional) tools that are used in education are 
\tool{PMD}~\cite{1620} and \tool{CppCheck}~\cite{1621}. These tools can also be integrated
in Continuous Integration practices, such as \tool{SonarQube}~\cite{1720}.
We also noticed that selfmade tools often make use of external tools
for specific tasks. For example, \tool{PyTA} is a wrapper for \tool{pylint}~\cite{1617},
adding custom code checks and improved error messages targeted at students.
\new{\tool{Hyperstyle}~\cite{n305} uses static analysers for different languages (PMD for Java, Detekt for Kotlin, and linters for JavaScript and Python), from which checks suitable for students are selected, categorized, and presented together with a grade.}

Tools can be used to analyse large collections of student code (papers focussing on
analysing program quality), and to support students in learning (papers focussing
on a tool for instruction), and in some papers tools have a dual role: the authors conclude
that student programs contain many flaws (identified by some tool), and therefore
that tool could be used as an instructional aid~\cite{1621}. However, it remains unclear
whether these tools are suitable for educating novices,
which is addressed by several papers. Nutbrown and Higgins~\cite{1620} analyse differences between tool assessment
and human assessment, and investigate the usefulness of such tools.

We have found only \nrc{six} papers that discuss \code{course materials}. \tool{Refactory}~\cite{8} is a non-digital
card game to learn the principles of refactoring by resolving code smells.
Other work analyses the readability of example
programs in programming books~\cite{388}.

Several papers discuss \code{programming assignments}, for example, by presenting coding 
guidelines~\cite{1692}
and code readability best
practices for students~\cite{1581}.
Stegeman et al. have developed a rubric for assessing the code quality of student code
\cite{581}, which we described
in more detail in section~\ref{sec:terms}.
Nandigam et al.~\cite{1079} describe assignments in which students are instructed to
explore and improve open source projects by measuring quality and applying refactorings
where needed.
\new{Tempero and Tu~\cite{n16} use code review assignments to asses how students
understand the concept of `maintainability'.}

\nrc{Nine} papers discuss teaching about the \code{programming process}. \\Stoecklin
et al.~\cite{1729} describe an approach for teaching refactoring through a series
of incremental lessons.
Abid et al.~\cite{1433} present an experiment about the timing of refactoring in a
student project. Lu et al.~\cite{968} introduce `Continuous Inspection' of code quality
in an educational setting. 
Passier et al.~\cite{387} describe how students can build an elegant JavaScript application
step by step.
 

\new{Papers labelled with \code{teaching method} address multiple
instructional elements.
Izu et al.~\cite{n55} present a teaching resource, consisting of a
textual explanation, a set of refactoring rules, and exercises, to
help students with identifying code smells in conditional
statements, and refactoring this code.
Crespo et al.~\cite{Crespo21} focus on the concept of `technical debt', and
compare two different teaching methods: a penalisation (based on
SonarQube metrics) and a rewarding strategy (with the metrics shown
in a leaderbord). }

\subsubsection{Learning outcome}

\code{Program quality} is a major category in papers, studying the programs
that students write with respect to quality characteristics.
These programs are mostly
analysed automatically by a tool to identify code smells and calculate quality metrics.
Examples of such large-scale studies, often analysing thousands of programs, are
a study of PMD rule violations in Java programs ~\cite{453}, smells in Scratch programs~\cite{911},
and indicators of semantic style differences~\cite{1909}.
\new{Cristea et al. \cite{n213} combine Formal Concept Analysis with Pylint,
to detect issues with object-oriented design and too complex code.}
\new{Groeneveld et al.~\cite{n302} analyse the correlation between code quality
and creativity, finding preliminary evidence for a larger number of issues in
projects with high creativity.}
\new{Grotov et al.~\cite{n261} compare the coding style and complexity of Python programs
written as regular scripts to code written in Jupyter computational notebooks.}
\new{Ma and Tilevich~\cite{n311} describe a set of anti-patterns that may arise when
students move from Java to Python, but still write code in the
much more verbose Java-style.}

The majority of program quality studies employ quantitative, descriptive methods, but
others take a qualitative approach. 
Some studies administer a survey to let teachers or students assess example code,
for example to collect suggestions from expert programmers~\cite{1880}.
Andrade and Brunet~\cite{410} studied whether students were able to
give useful feedback on the quality of other students' code. 

A few papers study a specific phenomenon related to code quality, such as the
`unencapsulated collection' design smell~\cite{1912}.
Studies that assess student code quality by hand are more rare.
Some papers compare code assessment by experts with code analysed by tools.

\nrc{Thirteen} papers focus on student \code{behaviours} with regards to code quality. Gilson et al.~\cite{910}
observed how student Scrum teams deal with quality issues during a one-year project.
Sripada and Reddy~\cite{449} also study student activities related to quality
in multiple iterations of a development process.
\new{Senger et al.~\cite{n304} replicate an earlier study with more and larger student programs, in which they
run the static analyser FindBugs and study the
correlation between the warnings found and correctness or struggling.}

\nrc{Eleven} studies are on \code{perceptions} of teachers and students, of which \nrc{five} mention the
term `readability'.
Kirk et al.~\cite{n30} study high-school teachers' ideas and needs regarding code style and quality through interviews. 
Wiese et al.~\cite{1091} investigate how beginner programmers assess the style of
example programs, which they later replicate with a different student population~\cite{1442}.
Note that this is \new{one of two} replication studies we identified in our set of papers.
An ITiCSE working group studied the differences in perceptions of code quality between
developers, teachers, and students~\cite{927}.
Fleury~\cite{b17} conducted interviews with students asking them
to evaluate and compare the style of several Java programs.


All \nrc{three} studies about \code{concept understanding} use a (quasi-) experimental approach.
Hermans and Aivaloglou~\cite{623} study the effect of smells in Scratch code when
students do comprehension tasks.

A few methods were not in our main list, such as 
`educational design research' for iteratively designing a code quality rubric~\cite{581}. Recently, we observed the use of machine
learning techniques~\cite{1160}. 



\subsection{Languages (RQ4)}

\begin{figure}[bt]
	\begin{center}
	\includegraphics[width=0.8\columnwidth]{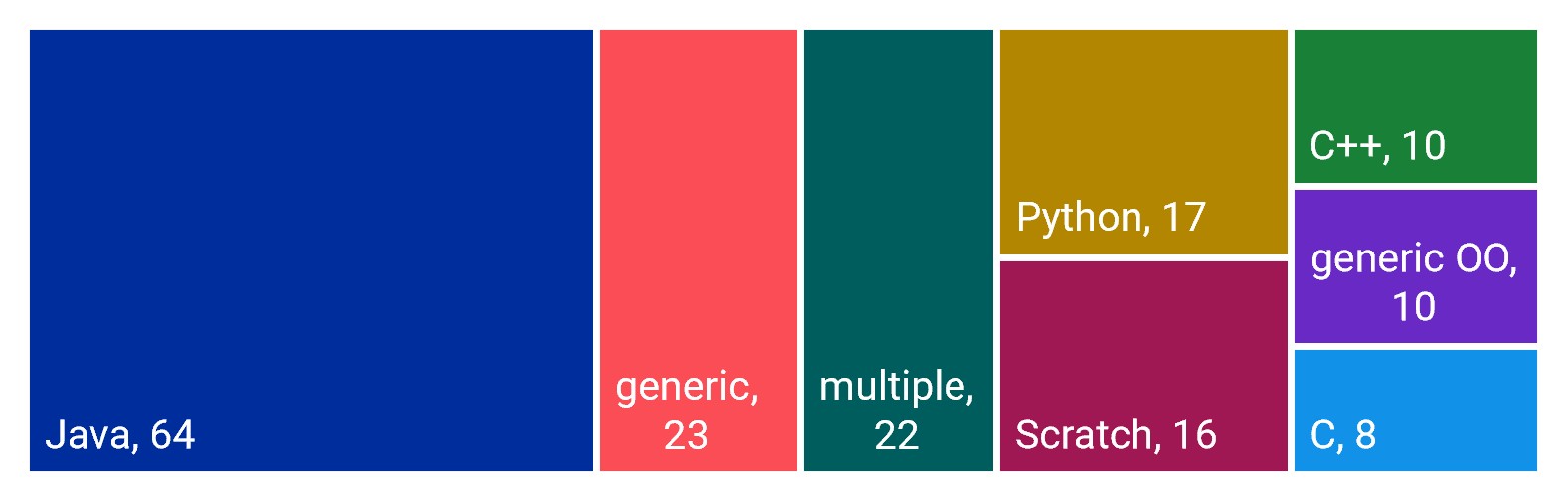}
	\end{center}
	\caption{Treemap of targeted programming languages. Languages with less
	than 5 papers are omitted.}
	\label{img:lang}
\end{figure}

Figure~\ref{img:lang} shows a treemap with the languages that are targeted
in the publications. Java and Python, popular general-purpose languages often
used in teaching, are the most prevalent text-based languages in this study. Less expected might
be the substantial number of papers dealing with programming in a block-based editor,
such as Scratch~\cite{1951,911} and Snap!~\cite{1790}.
These papers investigate code smells or present learning tools.
Other paradigms, such as functional programming, hardly appear.

\subsection{Trends (RQ5)}

\begin{figure}[bt]
	\begin{center}
	\includegraphics[width=0.9\columnwidth]{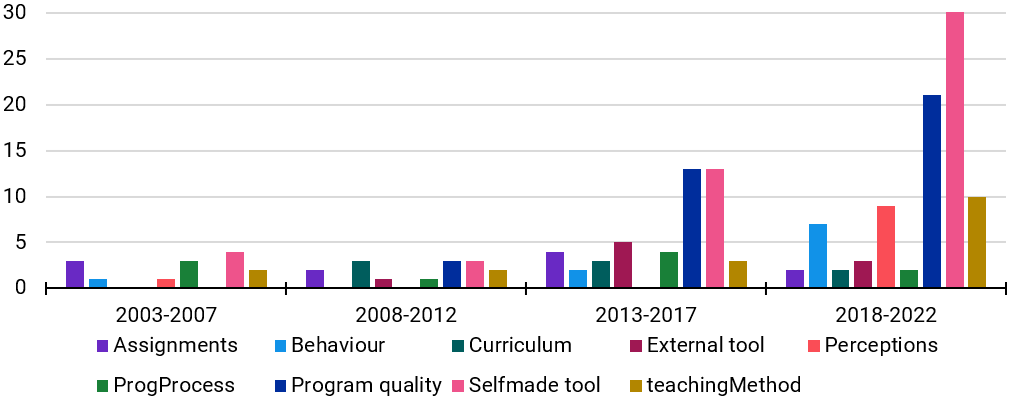}
	\end{center}
	\caption{Paper count per topic for the last 20 years, omitting topics with
	less than 8 papers.}
	\label{img:topic-trends}
\end{figure}

Figure~\ref{img:topic-trends} shows the trends with respect to paper topics.
We focus on the last twenty years, because we only have a small number of papers before that.
We notice that much of the program quality research appeared in the last decade. The number of papers on tools
has grown significantly, and the use of external tools is mostly a development from
the last ten years.
Studying perceptions is very much a recent development.

\subsection{Related fields (RQ6)}
\label{sec:relfields}

As discussed before, the term \textit{code quality} has no crystal clear definition.
During our search for relevant publications, we regularly came across papers
with a topic on the edges of our scope, as defined in section~\ref{sec:scope}.
In this section we list these topics, which are also relevant for learning and teaching
about code quality, and refer to 
literature reviews on these topics, if available.

\vspace{-7pt}
\paragraph{Software design education} While much of the research found in this study could be related
    to designing software, and in our definition we mention aspects such as decomposition
    and encapsulation, our mapping does not cover the broad field of teaching software
    design upfront. Instead, we focus on assessing the characteristics of the code after it has been
    written.

\vspace{-7pt}
  
\paragraph{Design patterns education} We included some papers dealing with design patterns,
    because they were used as a means to refactor existing code. There are several other
    papers that focus on teaching and learning of design patterns in
    general. 
 
 \vspace{-7pt}
    
\paragraph{Object-orientated programming} Abstraction, decomposition, and encapsulation are prominent topics
    in learning object-orientation,
     and contribute greatly to the quality of design and code.

\vspace{-6pt}
    
\paragraph{Interventions leading to improved code quality} Some interventions may lead
    to improved code quality, but are not specifically about code quality. Examples
    are pair programming
    , test-driven development~\cite{scatalon2019software}, 
    and peer review~\cite{indriasari2020review}. 
 
 \vspace{-6pt}
   
\paragraph{Computational thinking} Abstraction, decomposition, and modularization are important
    aspects of computational thinking~\cite{hsu2018learn}. 
    
 \vspace{-6pt}
     
\paragraph{Code similarity and plagiarism} Rather the reverse of assessing the various ways a program
    can be written, several studies focus on code similarity, 
    code clustering, 
    and detecting plagiarism~\cite{novak2019source}. 
 
 \vspace{-6pt}   
    
\paragraph{Program comprehension} This topic deals with the cognitive processes that programmers
apply when trying to understand programs~\cite{schulte2010introduction}. 
   
 
 \vspace{-6pt}
    
\paragraph{Automated assessment} Many systems for automated assessment of student programs
incorporate some kind of style feedback~\cite{paiva2022automated}.

\subsection{Threats to validity}

It is non-trivial to categorise a paper by its main method and topic. By only
assigning one label, we might miss some additional topics and methods. Aspects
were identified by looking for specific terms in the title and abstract; this
simplified method might not correctly represent an article's main focus.

\new{Although we performed an extensive database search followed
by snowballing, we might have discarded relevant work based on
an unclear title or abstract.}
Also, we have only included papers with code quality topics as their main focus.
Because code quality can be integrated in software engineering courses, and is an
element of overall software quality, we might miss some relevant research.

\section{Conclusion}
\label{sec:conclusion}

One of the earliest papers identified in this study on teaching programming style
concludes with `Perhaps the more recent structured languages such as PASCAL and C
will make some of this emphasis less critical'~\cite{1830}. Although tools and new languages
simplify implementing good coding style, the author has not foreseen the ongoing
issue with writing high-quality code.

We have conducted a systematic mapping study of code quality in education, which is the first overarching study on this topic. We identified and categorised \nrc{195} papers, studying paper characteristics, topics, domain-specific aspects, methods, and trends.
Code quality is an upcoming topic with an increasing number of studies.
Papers are published in a wide variety of venues on various topics.
Its main focus has been on developing and evaluating tools for feedback on code smells,
and suggestions for improvements and refactorings. Professional quality tools are
increasingly being used in (and adapted for) education.
Another major area is quality analysis
of student code.
We also observe that a growing number of studies target block-based
programming environments, emphasising the need to start early with this topic.
We have given several examples
of the diversity in research, and shown related fields.

Because the goal of a mapping study is to give a broad overview, a possible direction
for future work is to conduct a more in-depth literature study of a specific topic or aspect
identified in this study.
We would also encourage researchers to perform studies on the topics that have received
little attention so far, such as integrating code quality into the computing curricula,
developing and evaluating course materials, and studying student perceptions and behaviours.

\newpage \clearpage 

\bibliographystyle{ACM-Reference-Format}

\bibliography{full}

\newpage \clearpage 

\onecolumn

\renewcommand{\arraystretch}{1.8}
\setlength\tabcolsep{4pt}
\footnotesize\sffamily
\begin{longtable}{p{8cm}llll cccccc ccc}
\caption{Complete list of papers and coding.\\ CS=code smells, DP=design patterns, MT=metrics, RD=readability, RF=refactoring, SA=static analysis} \\ 
\label{table:full}  
\textbf{Title} & \textbf{Year} & \textbf{Topic} & \textbf{Method} & \textbf{Language} & \textbf{CS} & \textbf{DP} & \textbf{MT} & \textbf{RD} & \textbf{RF} & \textbf{SA} \\ \hline \endhead

"I know it when i see it" - Perceptions of code quality,  \citeauthor{927} \cite{927} & 2018 & Perceptions & DescrCorr & generic &  &  &  & rd &  &  \\ 
A card game for learning software-refactoring principles,  \citeauthor{8} \cite{8} & 2019 & Materials & Survey & generic-OO & cs &  &  &  & rf &  \\ 
A case study of the analysis of novice student programs,  \citeauthor{10} \cite{10} & 1999 & ProgramQuality & DescrCorr & C++ &  &  & mt &  &  & sa \\ 
A case study of the static analysis of the quality of novice student programs,  \citeauthor{11} \cite{11} & 1999 & ProgramQuality & DescrCorr & C++ &  &  & mt &  &  & sa \\ 
A critic for LISP,  \citeauthor{fischer1987} \cite{fischer1987} & 1987 & SelfmadeTool & Tool & Lisp &  &  &  &  &  &  \\ 
A design quality learning unit in OO modeling bridging the engineer and the artist,  \citeauthor{34} \cite{34} & 2011 & Curriculum & NoneUnclear & generic-OO &  &  & mt &  &  &  \\ 
A diagnosis system of programming styles using program patterns,  \citeauthor{37} \cite{37} & 2000 & SelfmadeTool & (Q)Experim & C &  &  &  & rd &  &  \\ 
A framework for the assessment and training of software refactoring competences,  \citeauthor{47} \cite{47} & 2019 & Curriculum & CaseStudy & generic &  &  &  &  & rf &  \\ 
A Heuristic Tool for Measuring Software Quality Using Program Language Standards,  \citeauthor{n153} \cite{n153} & 2022 & SelfmadeTool & Tool & Java &  &  &  &  &  &  \\ 
A Large-Scale Comparison of Python Code in Jupyter Notebooks and Scripts,  \citeauthor{n261} \cite{n261} & 2022 & ProgramQuality & DescrCorr & Python &  &  &  &  &  &  \\ 
A method for detecting bad smells and ITS application to software engineering education,  \citeauthor{59} \cite{59} & 2014 & SelfmadeTool & Tool & Java & cs &  &  &  & rf &  \\ 
A pedagogical approach in interleaving software quality concerns at an artificial intelligence course,  \citeauthor{n107} \cite{n107} & 2022 & TeachingMethod & NoneUnclear & Java &  &  & mt &  &  &  \\ 
A programming style taxonomy,  \citeauthor{92} \cite{92} & 1991 & Assignments & Other & generic &  &  &  &  &  &  \\ 
A Proposal of Coding Rule Learning Function in Java Programming Learning Assistant System,  \citeauthor{94} \cite{94} & 2016 & SelfmadeTool & Survey & Java &  &  &  & rd &  & sa \\ 
A Resource to Support Novices Refactoring Conditional Statements,  \citeauthor{n55} \cite{n55} & 2022 & TeachingMethod & (Q)Experim & C & cs &  &  & rd & rf &  \\ 
A study of loop style and abstraction in pedagogic practice,  \citeauthor{125} \cite{125} & 2011 & ProgramQuality & Qualitative & mutiple &  &  &  &  &  &  \\ 
A study on the quality mindedness of students,  \citeauthor{n217} \cite{n217} & 2022 & Perceptions & Survey & generic &  &  &  &  &  &  \\ 
A tool for diagnosing the quality of java program and a method for its effective utilization in education,  \citeauthor{149} \cite{149} & 2010 & ExternalTool & (Q)Experim & Java &  &  &  & rd &  &  \\ 
A Tutoring System to Learn Code Refactoring,  \citeauthor{n297} \cite{n297} & 2021 & SelfmadeTool & Tool & Java &  &  &  &  & rf &  \\ 
Academic coding guideline model - OCG,  \citeauthor{169} \cite{169} & 2014 & Assignments & Discussion & C &  &  &  & rd &  &  \\ 
Amelioration of Teaching Strategies by Exploring Code Quality and Submission Behavior,  \citeauthor{197} \cite{197} & 2019 & Behaviour & DescrCorr & C++ & cs &  &  &  &  &  \\ 
An Agile classroom experience: Teaching TDD and refactoring,  \citeauthor{200} \cite{200} & 2008 & ProgProcess & Experience & Java &  &  &  &  & rf &  \\ 
An Applicability Study on Refactoring Principles in Reading-Based Programming Learning,  \citeauthor{n257} \cite{n257} & 2022 & Materials & (Q)Experim & Java &  & dp &  &  & rf &  \\ 
An automated assessment system for analysis of coding convention violations in Java programming assignments,  \citeauthor{212} \cite{212} & 2018 & SelfmadeTool & Tool & Java &  &  &  & rd &  &  \\ 
An Automatic Grading System for a High School-level Computational Thinking Course,  \citeauthor{n263} \cite{n263} & 2022 & SelfmadeTool & QuantOther & Haskell &  &  &  &  &  &  \\ 
An empirical study of COBOL programs via a style analyzer: The benefits of good programming style,  \citeauthor{224} \cite{224} & 1989 & Behaviour & DescrCorr & Cobol &  &  & mt &  &  &  \\ 
An empirical study of iterative improvement in programming assignments,  \citeauthor{225} \cite{225} & 2015 & Behaviour & DescrCorr & C++ &  &  & mt &  &  &  \\ 
An empirical study on students' ability to comprehend design patterns,  \citeauthor{232} \cite{232} & 2008 & ConceptUnd & (Q)Experim & mutiple &  & dp & mt &  &  &  \\ 
An Incremental Model for Developing Educational Critiquing Systems: Experiences with the Java Critiquer,  \citeauthor{b3} \cite{b3} & 2008 & SelfmadeTool & DescrCorr & Java &  &  &  &  &  &  \\ 
An innovative approach to teaching refactoring,  \citeauthor{266} \cite{266} & 2006 & ProgProcess & NoneUnclear & generic-OO &  &  &  &  & rf &  \\ 
An instructional aid for student programs,  \citeauthor{267} \cite{267} & 1980 & SelfmadeTool & DescrCorr & Fortran &  &  &  &  &  &  \\ 
Analysis of Learning Behavior in an Automated Programming Assessment Environment: A Code Quality Perspective,  \citeauthor{292} \cite{292} & 2020 & Behaviour & QuantOther & Java &  &  &  &  &  &  \\ 
Analyzing students' software redesign strategies,  \citeauthor{300} \cite{300} & 2016 & Behaviour & Qualitative & Java &  & dp &  &  & rf &  \\ 
Anomaly Detection for Early Warning in Object-oriented Programming Course,  \citeauthor{n319} \cite{n319} & 2021 & SelfmadeTool & Tool & Java &  &  &  &  &  & sa \\ 
Applying Code Quality Detection in Online Programming Judge,  \citeauthor{317} \cite{317} & 2020 & ExternalTool & DescrCorr & Python &  &  &  &  &  &  \\ 
Applying gamification to motivate students to write high-quality code in programming assignments,  \citeauthor{318} \cite{318} & 2019 & TeachingMethod & (Q)Experim & C &  &  & mt &  &  &  \\ 
Are Undergraduate Creative Coders Clean Coders? A Correlation Study,  \citeauthor{n302} \cite{n302} & 2022 & ProgramQuality & DescrCorr & Java &  &  &  &  &  &  \\ 
ASPA: A Static Analyser to Support Learning and Continuous Feedback on Programming Courses. An Empirical Validation,  \citeauthor{n275} \cite{n275} & 2022 & SelfmadeTool & Survey & Python &  &  &  &  &  & sa \\ 
Assessing software quality of agile student projects by data-mining software repositories,  \citeauthor{349} \cite{349} & 2019 & ProgramQuality & DescrCorr & unknown &  &  & mt &  &  &  \\ 
Assessing the quality of programs: A topic for the CS2 course,  \citeauthor{sanders1987} \cite{sanders1987} & 1987 & ProgramQuality & Discussion & generic &  &  &  &  &  &  \\ 
Assessing Understanding of Maintainability using Code Review,  \citeauthor{n16} \cite{n16} & 2021 & Assignments & (Q)Experim & generic &  &  &  &  &  &  \\ 
Automated critique of early programming antipatterns,  \citeauthor{366} \cite{366} & 2019 & SelfmadeTool & Tool & Java &  &  &  &  &  & sa \\ 
Automatic analysis of functional program style,  \citeauthor{michaelson1996} \cite{michaelson1996} & 1996 & SelfmadeTool & Tool & SML &  &  &  &  &  &  \\ 
Automatic assessment aids for Pascal programs,  \citeauthor{rees1982} \cite{rees1982} & 1982 & SelfmadeTool & DescrCorr & Pascal &  &  &  &  &  &  \\ 
Automatic Assessment of the Design Quality of Student Python and Java Programs,  \citeauthor{n178} \cite{n178} & 2022 & SelfmadeTool & (Q)Experim & mutiple &  &  &  & rd &  &  \\ 
Automatic detection of bad programming habits in scratch,  \citeauthor{moreno2014} \cite{moreno2014} & 2014 & ProgramQuality & DescrCorr & Scratch &  &  &  &  &  &  \\ 
Automatic grader for programming assignment using source code analyzer,  \citeauthor{375} \cite{375} & 2014 & SelfmadeTool & DescrCorr & mutiple &  &  &  &  &  &  \\ 
Automatic programming assessment,  \citeauthor{377} \cite{377} & 1993 & ProgramQuality & DescrCorr & Pascal &  &  & mt &  &  &  \\ 
AutoStyle: Toward coding style feedback at scale,  \citeauthor{386} \cite{386} & 2015 & SelfmadeTool & Tool & mutiple &  &  &  &  &  &  \\ 
Bad Smells in Scratch Projects: A Preliminary Analysis,  \citeauthor{vargas2019} \cite{vargas2019} & 2019 & ProgramQuality & DescrCorr & Scratch & cs &  &  &  &  &  \\ 
Beautiful JavaScript: How to guide students to create good and elegant code,  \citeauthor{387} \cite{387} & 2014 & ProgProcess & NoneUnclear & JavaScript &  &  &  &  & rf &  \\ 
Beauty and the Beast: on the readability of object-oriented example programs,  \citeauthor{388} \cite{388} & 2016 & Materials & DescrCorr & Java &  &  &  & rd &  &  \\ 
But my program runs! Discourse rules for novice programmers,  \citeauthor{Joni86} \cite{Joni86} & 1986 & Assignments & NoneUnclear & Pascal &  &  &  &  &  &  \\ 
Can students help themselves? An investigation of students' feedback on the quality of the source code,  \citeauthor{410} \cite{410} & 2019 & ProgramQuality & Survey & Python &  &  &  &  &  &  \\ 
Carrot and Stick approaches revisited when managing Technical Debt in an educational context,  \citeauthor{Crespo21} \cite{Crespo21} & 2021 & TeachingMethod & (Q)Experim & Java &  &  & mt &  &  &  \\ 
Challenges of knowledge component modeling: A software engineering case study,  \citeauthor{n161} \cite{n161} & 2022 & Materials & CaseStudy & generic &  &  &  &  & rf &  \\ 
Clean Code - Delivering A Lightweight Course,  \citeauthor{n243} \cite{n243} & 2021 & TeachingMethod & NoneUnclear & Java &  &  &  &  &  &  \\ 
Clean Code and Design Educational Tool,  \citeauthor{n1} \cite{n1} & 2021 & SelfmadeTool & Tool & C\# & cs &  &  & rd &  &  \\ 
Clean Code Tutoring: Makings of a Foundation,  \citeauthor{n155} \cite{n155} & 2022 & SelfmadeTool & (Q)Experim & C\# &  &  &  & rd & rf &  \\ 
Cleangame: Gamifying the identification of code smells,  \citeauthor{442} \cite{442} & 2019 & SelfmadeTool & (Q)Experim & Java & cs &  &  &  & rf &  \\ 
Code Comprehension Activities in Undergraduate Software Engineering Course - A Case Study,  \citeauthor{449} \cite{449} & 2015 & Behaviour & DescrCorr & mutiple &  &  &  &  & rf &  \\ 
Code Perfumes: Reporting Good Code to Encourage Learners,  \citeauthor{n18} \cite{n18} & 2021 & ProgramQuality & DescrCorr & Scratch & cs &  &  &  &  &  \\ 
Code Quality Defects Across Introductory Programming Topics,  \citeauthor{n303} \cite{n303} & 2022 & ProgramQuality & DescrCorr & Python &  &  &  &  &  &  \\ 
Code Quality Improvement for All: Automated Refactoring for Scratch,  \citeauthor{452} \cite{452} & 2019 & SelfmadeTool & (Q)Experim & Scratch & cs &  & mt &  & rf &  \\ 
Code quality issues in student programs,  \citeauthor{453} \cite{453} & 2017 & ProgramQuality & DescrCorr & Java &  &  &  &  &  &  \\ 
Code quality: Examining the efficacy of automated tools,  \citeauthor{454} \cite{454} & 2017 & ExternalTool & DescrCorr & Python &  &  & mt &  &  &  \\ 
CompareCFG: Providing Visual Feedback on Code Quality Using Control Flow Graphs,  \citeauthor{487} \cite{487} & 2020 & SelfmadeTool & Tool & Java &  &  &  &  &  &  \\ 
Comparison of software quality in the work of children and professional developers based on their classroom exercises,  \citeauthor{493} \cite{493} & 2015 & ProgramQuality & DescrCorr & Java &  &  &  &  &  &  \\ 
Comprehension and application of design patterns by novice software engineers,  \citeauthor{499} \cite{499} & 2018 & ConceptUnd & (Q)Experim & Java &  & dp &  &  &  &  \\ 
Dependency Analysis for Learning Class Structure for Novice Java Programmer,  \citeauthor{558} \cite{558} & 2011 & TeachingMethod & CaseStudy & Java &  &  &  & rd & rf &  \\ 
Design of e-activities for the learning of code refactoring tasks,  \citeauthor{571} \cite{571} & 2014 & TeachingMethod & Experience & generic &  &  &  &  & rf &  \\ 
Design patterns in scientific software,  \citeauthor{575} \cite{575} & 2004 & TeachingMethod & NoneUnclear & Java &  & dp &  &  & rf &  \\ 
Designing a Programming Game to Improve Children's Procedural Abstraction Skills in Scratch,  \citeauthor{580} \cite{580} & 2020 & SelfmadeTool & (Q)Experim & Scratch & cs &  &  &  &  &  \\ 
Designing a rubric for feedback on code quality in programming courses,  \citeauthor{581} \cite{581} & 2016 & Assignments & Other & generic &  &  &  &  &  &  \\ 
Detecting and Addressing Design Smells in Novice Processing Programs,  \citeauthor{591} \cite{591} & 2019 & ProgramQuality & DescrCorr & Processing & cs &  &  &  & rf & sa \\ 
Development of a refactoring learning environment,  \citeauthor{605} \cite{605} & 2011 & SelfmadeTool & Tool & Java &  &  &  &  & rf &  \\ 
Do code smells hamper novice programming? A controlled experiment on Scratch programs,  \citeauthor{623} \cite{623} & 2016 & ConceptUnd & (Q)Experim & Scratch & cs &  &  &  &  &  \\ 
Documentation Standards for Beginning Students,  \citeauthor{628} \cite{628} & 1976 & Assignments & NoneUnclear & mutiple &  &  &  & rd &  &  \\ 
Does Static Analysis Help Software Engineering Students?,  \citeauthor{634} \cite{634} & 2020 & ExternalTool & DescrCorr & Java &  &  &  &  &  & sa \\ 
Dr. Scratch: Automatic analysis of scratch projects to assess and foster computational thinking,  \citeauthor{moreno2015dr} \cite{moreno2015dr} & 2015 & SelfmadeTool & (Q)Experim & Scratch &  &  &  &  &  &  \\ 
DrPython-WEB: A Tool to Help Teaching Well-Written Python Programs,  \citeauthor{n209} \cite{n209} & 2022 & SelfmadeTool & Tool & Python &  &  &  &  &  &  \\ 
Earthworm - Automated decomposition suggestions,  \citeauthor{647} \cite{647} & 2018 & SelfmadeTool & Tool & Python &  &  &  &  & rf & sa \\ 
Effectively teaching coding standards in programming,  \citeauthor{li2005} \cite{li2005} & 2005 & Perceptions & Survey & generic &  &  &  &  &  &  \\ 
Effects of technical debt awareness: A classroom study,  \citeauthor{tonin2017} \cite{tonin2017} & 2017 & TeachingMethod & Qualitative & generic &  &  &  &  &  &  \\ 
Encapsulation and Reuse as Viewed by Java Students,  \citeauthor{b17} \cite{b17} & 2001 & Perceptions & Qualitative & Java &  &  &  &  &  &  \\ 
Enhancing Abstraction in App Inventor with Generic Event Handlers,  \citeauthor{patton2019} \cite{patton2019} & 2019 & SelfmadeTool & Tool & APPInventor & cs &  &  &  & rf &  \\ 
Enhancing block-based programming pedagogy to promote the culture of quality from the ground up - a position paper,  \citeauthor{712} \cite{712} & 2017 & Curriculum & Discussion & generic-block &  &  &  &  &  &  \\ 
Evaluating Code Improvements in Software Quality Course Projects,  \citeauthor{n24} \cite{n24} & 2022 & TeachingMethod & (Q)Experim & Java &  &  &  &  &  & sa \\ 
Evolving an integrated curriculum for object-oriented analysis and design,  \citeauthor{ramnath2008} \cite{ramnath2008} & 2008 & Curriculum & Experience & generic-OO &  & dp &  &  & rf &  \\ 
Exploration of Experimental Teaching Reforms on C Programming Design Course,  \citeauthor{n255} \cite{n255} & 2021 & TeachingMethod & DescrCorr & C &  &  &  &  &  &  \\ 
Exploring Metrics for the Analysis of Code Submissions in an Introductory Data Science Course,  \citeauthor{Nguyen21} \cite{Nguyen21} & 2021 & ProgramQuality & DescrCorr & Python &  &  & mt &  &  &  \\ 
Five reasons for including technical debt in the software engineering curriculum,  \citeauthor{falessi2015} \cite{falessi2015} & 2015 & Curriculum & NoneUnclear & generic &  &  &  &  &  &  \\ 
Foobaz: Variable name feedback for student code at scale,  \citeauthor{824} \cite{824} & 2015 & SelfmadeTool & Survey & Python &  &  &  &  &  &  \\ 
Forming groups for collaborative learning in introductory computer programming courses based on students' programming styles: An empirical study,  \citeauthor{829} \cite{829} & 2006 & ProgProcess & (Q)Experim & C &  &  & mt &  &  &  \\ 
Fostering the comprehension of the object-oriented programming paradigm by a virtual lab exercise,  \citeauthor{thurner2019} \cite{thurner2019} & 2019 & TeachingMethod & Experience & Java &  &  &  &  &  &  \\ 
FrenchPress gives students automated feedback on Java program flaws,  \citeauthor{835} \cite{835} & 2015 & SelfmadeTool & Survey & Java &  &  &  &  &  &  \\ 
Function Names: Quantifying the Relationship Between Identifiers and Their Functionality to Improve Them,  \citeauthor{n337} \cite{n337} & 2022 & SelfmadeTool & Tool & Java &  &  &  & rd &  &  \\ 
Gamification based learning environment for computer science students,  \citeauthor{853} \cite{853} & 2020 & SelfmadeTool & Tool & mutiple &  &  &  &  &  & sa \\ 
Helping Student Programmers Through Industrial-Strength Static Analysis: A Replication Study,  \citeauthor{n304} \cite{n304} & 2022 & Behaviour & DescrCorr & Java &  &  &  &  &  & sa \\ 
High School Teachers' Understanding of Code Style,  \citeauthor{893} \cite{893} & 2020 & Perceptions & Qualitative & generic &  &  &  &  &  &  \\ 
How junior developers deal with their technical debt?,  \citeauthor{910} \cite{910} & 2020 & Behaviour & Mixed & mutiple &  &  &  &  &  & sa \\ 
How kids code and how we know: An exploratory study on the scratch repository,  \citeauthor{911} \cite{911} & 2016 & ProgramQuality & DescrCorr & Scratch & cs &  &  &  &  &  \\ 
How teachers would help students to improve their code,  \citeauthor{917} \cite{917} & 2019 & ProgramQuality & Survey & generic &  &  &  &  &  &  \\ 
How to improve code quality by measurement and refactoring,  \citeauthor{918} \cite{918} & 2016 & ProgProcess & DescrCorr & Java &  &  & mt &  & rf & sa \\ 
Human vs. Automated coding style grading in computing education,  \citeauthor{924} \cite{924} & 2019 & ProgramQuality & DescrCorr & C++ &  &  &  &  &  & sa \\ 
Hyperstyle: A Tool for Assessing the Code Quality of Solutions to Programming Assignments,  \citeauthor{n305} \cite{n305} & 2022 & SelfmadeTool & Tool & mutiple &  &  &  & rd & rf &  \\ 
Impact of aspect-oriented programming on the quality of novices' programs: A comparative study,  \citeauthor{939} \cite{939} & 2013 & ProgramQuality & (Q)Experim & C\# &  &  & mt & rd &  &  \\ 
Implementing a set of guidelines for CS majors in the production of program code,  \citeauthor{poole1996} \cite{poole1996} & 1996 & Assignments & Survey & Modula2 &  &  &  &  &  &  \\ 
Improving Feedback on GitHub Pull Requests: A Bots Approach,  \citeauthor{957} \cite{957} & 2019 & SelfmadeTool & Mixed & generic-OO & cs &  &  &  &  & sa \\ 
Improving Readability of Scratch Programs with Search-based Refactoring,  \citeauthor{n272} \cite{n272} & 2021 & SelfmadeTool & Tool & Scratch &  &  &  & rd & rf &  \\ 
Improving students programming quality with the continuous inspection process: a social coding perspective,  \citeauthor{968} \cite{968} & 2019 & ProgProcess & (Q)Experim & Java &  &  &  &  &  &  \\ 
Improving the software quality - An educational approach,  \citeauthor{974} \cite{974} & 2017 & SelfmadeTool & Tool & C\# &  & dp &  &  & rf &  \\ 
Integrating Antipatterns into the Computer Science Curriculum,  \citeauthor{1005} \cite{1005} & 2009 & Curriculum & NoneUnclear & generic-OO &  & dp &  &  & rf &  \\ 
Investigating code quality tools in the context of software engineering education,  \citeauthor{1028} \cite{1028} & 2017 & ExternalTool & DescrCorr & Java &  &  & mt &  & rf &  \\ 
Investigating static analysis errors in student Java programs,  \citeauthor{1032} \cite{1032} & 2017 & ProgramQuality & DescrCorr & Java &  &  &  &  &  & sa \\ 
Investigation of the relationship between program correctness and programming style,  \citeauthor{1041} \cite{1041} & 1999 & ProgramQuality & DescrCorr & mutiple &  &  & mt & rd &  &  \\ 
Japroch: A tool for checking programming style,  \citeauthor{makela2004} \cite{makela2004} & 2004 & SelfmadeTool & Tool & Java &  &  &  &  &  &  \\ 
JMetricGrader: A software for evaluating student projects using design object-oriented metrics and neural networks,  \citeauthor{n8} \cite{n8} & 2022 & ProgramQuality & QuantOther & Java &  &  & mt &  &  &  \\ 
Learning appreciation for design patterns by doing it the hard way first,  \citeauthor{skrien2003} \cite{skrien2003} & 2003 & TeachingMethod & Experience & Java &  & dp &  &  & rf &  \\ 
Learning software engineering principles using open source software,  \citeauthor{1079} \cite{1079} & 2008 & Assignments & NoneUnclear & Java &  &  & mt & rd & rf &  \\ 
Learning to listen for design,  \citeauthor{1083} \cite{1083} & 2019 & ProgProcess & Discussion & generic & cs & dp &  &  & rf &  \\ 
Linking code readability, structure, and comprehension among novices: It's complicated,  \citeauthor{1091} \cite{1091} & 2019 & Perceptions & Survey & mutiple &  &  &  & rd &  &  \\ 
Litterbox: A linter for scratch programs,  \citeauthor{Fraser21} \cite{Fraser21} & 2021 & SelfmadeTool & Tool & Scratch & cs &  &  &  &  &  \\ 
Measuring static quality of student code,  \citeauthor{1121} \cite{1121} & 2011 & ProgramQuality & DescrCorr & Java &  &  & mt &  &  &  \\ 
Measuring students' source code quality in software development projects through commit-impact analysis,  \citeauthor{Hamer21} \cite{Hamer21} & 2021 & Behaviour & DescrCorr & mutiple &  &  & mt &  &  &  \\ 
Mind the Gap: Searching for Clarity in NCEA,  \citeauthor{n58} \cite{n58} & 2021 & Materials & Mixed & generic &  &  &  &  &  &  \\ 
Mining student CVS repositories for performance indicators,  \citeauthor{1145} \cite{1145} & 2005 & Behaviour & DescrCorr & mutiple &  &  &  &  &  &  \\ 
Modeling Learners Programming Skills and Question Levels Through Machine Learning,  \citeauthor{1160} \cite{1160} & 2020 & ProgramQuality & QuantOther & mutiple &  &  &  & rd &  &  \\ 
Novice Programmers and Software Quality: Trends and Implications,  \citeauthor{1190} \cite{1190} & 2017 & ProgramQuality & DescrCorr & Scratch & cs &  &  &  &  &  \\ 
On assuring learning about code quality,  \citeauthor{1201} \cite{1201} & 2020 & Curriculum & CaseStudy & generic &  &  &  &  &  &  \\ 
On the Use of FCA Models in Static Analysis Tools to Detect Common Errors in Programming,  \citeauthor{n213} \cite{n213} & 2021 & ProgramQuality & DescrCorr & Python &  &  &  &  &  & sa \\ 
Pirate plunder: Game-based computational thinking using scratch blocks,  \citeauthor{1268} \cite{1268} & 2018 & SelfmadeTool & Tool & Scratch & cs &  &  &  &  &  \\ 
Program decomposition and complexity in CS1,  \citeauthor{1310} \cite{1310} & 2015 & TeachingMethod & (Q)Experim & C &  &  & mt &  &  &  \\ 
Programming style in introductory programming courses,  \citeauthor{1320} \cite{1320} & 2015 & Curriculum & NoneUnclear & generic &  &  &  &  &  &  \\ 
Promoting Code Quality via Automated Feedback on Student Submissions,  \citeauthor{n267} \cite{n267} & 2021 & SelfmadeTool & Tool & mutiple &  &  &  &  &  &  \\ 
Qualitative aspects of students' programs: Can we make them measurable?,  \citeauthor{1337} \cite{1337} & 2016 & SelfmadeTool & (Q)Experim & Python &  &  &  &  &  &  \\ 
Quality Assessment of Learners' Programs by Grouping Source Code Metrics,  \citeauthor{n156} \cite{n156} & 2021 & ProgramQuality & QuantOther & Lua &  &  & mt &  &  &  \\ 
Readable vs. Writable Code: A Survey of Intermediate Students' Structure Choices,  \citeauthor{n347} \cite{n347} & 2022 & Perceptions & Survey & Java &  &  &  & rd &  &  \\ 
RefacTutor: An Interactive Tutoring System for Software Refactoring,  \citeauthor{1432} \cite{1432} & 2020 & SelfmadeTool & Tool & Java &  &  &  &  & rf &  \\ 
Reflections on teaching refactoring: A tale of two projects,  \citeauthor{1433} \cite{1433} & 2015 & ProgProcess & (Q)Experim & Java &  &  &  &  &  &  \\ 
ReLE - a refactoring supporting tool,  \citeauthor{1437} \cite{1437} & 2011 & SelfmadeTool & Tool & Java &  &  &  &  & rf &  \\ 
Replicating novices' struggles with coding style,  \citeauthor{1442} \cite{1442} & 2019 & Perceptions & Survey & mutiple &  &  &  & rd &  &  \\ 
Research and practice on education of SQA at source code level,  \citeauthor{1449} \cite{1449} & 2011 & TeachingMethod & CaseStudy & generic &  &  &  &  &  &  \\ 
Salient elements in novice solutions to code writing problems,  \citeauthor{1478} \cite{1478} & 2011 & ProgramQuality & Qualitative & mutiple &  &  &  &  &  &  \\ 
Scale-driven automatic hint generation for coding style,  \citeauthor{1479} \cite{1479} & 2016 & SelfmadeTool & (Q)Experim & mutiple &  &  &  &  &  &  \\ 
Serious refactoring games,  \citeauthor{haendler2019} \cite{haendler2019} & 2019 & Materials & NoneUnclear & generic-OO & cs &  &  &  & rf &  \\ 
Smells in block-based programming languages,  \citeauthor{1517} \cite{1517} & 2016 & ProgramQuality & DescrCorr & generic-block & cs &  &  &  &  &  \\ 
Software analytics to support students in object-oriented programming tasks: an empirical study,  \citeauthor{ardimento2020} \cite{ardimento2020} & 2020 & ProgramQuality & (Q)Experim & Java &  &  &  &  &  &  \\ 
Software clones in scratch projects: On the presence of copy-and-paste in computational thinking learning,  \citeauthor{robles2017} \cite{robles2017} & 2017 & ProgramQuality & DescrCorr & Scratch &  &  &  &  &  &  \\ 
Software engineer education support system ALECSS utilizing devOps tools,  \citeauthor{1532} \cite{1532} & 2016 & SelfmadeTool & DescrCorr & Java &  &  &  &  &  & sa \\ 
Software metrics as a programming training tool,  \citeauthor{1545} \cite{1545} & 1990 & Assignments & (Q)Experim & Cobol &  &  & mt &  &  &  \\ 
Software Quality as a Subsidy for Teaching Programming,  \citeauthor{n268} \cite{n268} & 2021 & TeachingMethod & DescrCorr & Java &  &  &  &  &  &  \\ 
Software Quality Metrics Calculations for Java Programming Learning Assistant System,  \citeauthor{1574} \cite{1574} & 2020 & SelfmadeTool & DescrCorr & Java &  &  & mt &  &  &  \\ 
Software readability practices and the importance of their teaching,  \citeauthor{1581} \cite{1581} & 2016 & Assignments & Survey & generic-OO & cs &  &  & rd &  &  \\ 
Sprinter: A Didactic Linter for Structured Programming,  \citeauthor{Alfredo22} \cite{Alfredo22} & 2022 & SelfmadeTool & Tool & Java &  &  &  &  &  &  \\ 
Static analyses in python programming courses,  \citeauthor{1617} \cite{1617} & 2019 & SelfmadeTool & (Q)Experim & Python &  &  &  &  &  & sa \\ 
Static analysis of programming exercises: Fairness, usefulness and a method for application,  \citeauthor{1620} \cite{1620} & 2016 & ExternalTool & DescrCorr & Java &  &  &  &  &  & sa \\ 
Static analysis of source code written by novice programmers,  \citeauthor{1621} \cite{1621} & 2017 & ExternalTool & DescrCorr & C &  &  &  &  &  & sa \\ 
Static Analysis of Students' Java Programs,  \citeauthor{1622} \cite{1622} & 2004 & SelfmadeTool & Tool & Java &  &  & mt &  &  & sa \\ 
Structural analysis of source code collected from programming contests,  \citeauthor{1632} \cite{1632} & 2014 & ProgramQuality & QuantOther & C++ &  &  &  &  &  &  \\ 
Student Refactoring Behaviour in a Programming Tutor,  \citeauthor{1643} \cite{1643} & 2020 & Behaviour & DescrCorr & Java &  &  &  &  & rf &  \\ 
Students Projects' Source Code Changes Impact on Software Quality Through Static Analysis,  \citeauthor{n93} \cite{n93} & 2021 & Behaviour & DescrCorr & mutiple &  &  &  &  &  & sa \\ 
Studying Software Metrics Based on Real-World Software Systems,  \citeauthor{1658} \cite{1658} & 2007 & Assignments & NoneUnclear & generic &  &  & mt &  &  &  \\ 
Supporting Students in C++ Programming Courses with Automatic Program Style Assessment,  \citeauthor{b42} \cite{b42} & 2004 & SelfmadeTool & Qualitative & C++ &  &  &  &  &  &  \\ 
Teacher Mate: A Support Tool for Teaching Code Quality,  \citeauthor{1675} \cite{1675} & 2020 & SelfmadeTool & DescrCorr & Java &  &  &  &  &  &  \\ 
Teaching code quality in high school programming courses - Understanding teachers' needs,  \citeauthor{n30} \cite{n30} & 2022 & Perceptions & Qualitative & generic &  &  &  &  &  &  \\ 
Teaching Defensive Programming in Java,  \citeauthor{1692} \cite{1692} & 2004 & Assignments & Survey & Java &  &  &  &  &  &  \\ 
Teaching design patterns using a family of games,  \citeauthor{1693} \cite{1693} & 2009 & Assignments & Experience & Java &  & dp &  &  & rf &  \\ 
Teaching programming style with ugly code,  \citeauthor{1704} \cite{1704} & 2013 & SelfmadeTool & Tool & Java &  &  &  & rd &  &  \\ 
Teaching software quality via source code inspection tool,  \citeauthor{1720} \cite{1720} & 2017 & SelfmadeTool & (Q)Experim & mutiple &  &  &  &  &  &  \\ 
Teaching students to build well formed object-oriented methods through refactoring,  \citeauthor{1729} \cite{1729} & 2007 & ProgProcess & NoneUnclear & generic-OO &  &  &  &  & rf &  \\ 
Teaching students to recognize and implement good coding style,  \citeauthor{1730} \cite{1730} & 2017 & ProgramQuality & (Q)Experim & Python &  &  &  &  &  &  \\ 
Teaching the culture of quality from the ground up: Novice-tailored quality improvement for scratch programmers,  \citeauthor{tilevich2020} \cite{tilevich2020} & 2020 & SelfmadeTool & Mixed & Scratch &  &  &  &  & rf &  \\ 
The effect of reporting Known issues on students' work,  \citeauthor{1782} \cite{1782} & 2018 & Perceptions & (Q)Experim & C++ &  &  &  &  &  &  \\ 
The Five Million Piece Puzzle: Finding Answers in 500,000 Snap-Projects,  \citeauthor{1790} \cite{1790} & 2019 & ProgramQuality & DescrCorr & Snap! & cs &  &  &  &  &  \\ 
The impact of automated code quality feedback in programming education,  \citeauthor{1796} \cite{1796} & 2017 & ExternalTool & (Q)Experim & mutiple &  &  &  &  &  &  \\ 
The LAN-simulation: A refactoring teaching example,  \citeauthor{1807} \cite{1807} & 2005 & Assignments & Experience & Java &  &  &  &  & rf &  \\ 
The Role of Source Code Vocabulary in Programming Teaching and Learning,  \citeauthor{1824} \cite{1824} & 2020 & SelfmadeTool & (Q)Experim & Python &  &  &  & rd &  &  \\ 
The teaching of documentation and good programming style in a liberal arts computer science program,  \citeauthor{1830} \cite{1830} & 1980 & TeachingMethod & NoneUnclear & Basic &  &  &  &  &  &  \\ 
Tool assisted identifier naming for improved software readability: An empirical study,  \citeauthor{1854} \cite{1854} & 2005 & SelfmadeTool & (Q)Experim & Java &  &  &  & rd &  &  \\ 
Towards an empirically validated model for assessment of code quality,  \citeauthor{1875} \cite{1875} & 2014 & Assignments & Qualitative & generic &  &  &  &  &  &  \\ 
Towards generalizing expert programmers' suggestions for novice programmers,  \citeauthor{1880} \cite{1880} & 2013 & ProgramQuality & Survey & Alice-LG &  &  &  &  &  & sa \\ 
Understanding recurring quality problems and their impact on code sharing in block-based software,  \citeauthor{1908} \cite{1908} & 2017 & ProgramQuality & DescrCorr & Scratch & cs &  &  &  &  &  \\ 
Understanding Refactoring Tasks over Time: A Study Using Refactoring Graphs,  \citeauthor{n88} \cite{n88} & 2022 & Behaviour & (Q)Experim & Java &  &  &  &  & rf &  \\ 
Understanding Semantic Style by Analysing Student Code,  \citeauthor{1909} \cite{1909} & 2018 & ProgramQuality & DescrCorr & Java &  &  &  &  &  &  \\ 
Unencapsulated collection - A teachable design smell,  \citeauthor{1912} \cite{1912} & 2018 & ProgramQuality & CaseStudy & generic-OO & cs &  &  &  & rf &  \\ 
Unreadable code in novice developers,  \citeauthor{Avila21} \cite{Avila21} & 2021 & Perceptions & Survey & generic &  &  &  & rd &  &  \\ 
Using examples as guideposts for programming exercises: Providing support while preserving the challenge,  \citeauthor{n126} \cite{n126} & 2021 & Assignments & CaseStudy & C++ &  &  &  &  & rf &  \\ 
Using pirate plunder to develop children's abstraction skills in scratch,  \citeauthor{1951} \cite{1951} & 2019 & SelfmadeTool & (Q)Experim & Scratch & cs &  &  &  &  &  \\ 
Using project-based approach to teach design patterns: An Experience Report,  \citeauthor{n33} \cite{n33} & 2021 & TeachingMethod & (Q)Experim & Java & cs & dp &  &  & rf &  \\ 
Using software metrics tools for maintenance decisions: a classroom exercise,  \citeauthor{1954} \cite{1954} & 1996 & ExternalTool & CaseStudy & unknown &  &  & mt &  &  &  \\ 
Using static analysis tools to assist student project evaluation,  \citeauthor{1956} \cite{1956} & 2020 & ExternalTool & DescrCorr & Python &  &  &  &  &  & sa \\ 
Using Verilog LOGISCOPE to analyze student programs,  \citeauthor{1964} \cite{1964} & 1998 & ExternalTool & DescrCorr & C++ &  &  &  &  &  & sa \\ 
Utilizing software engineering education support system ALECSS at an actual software development experiment: A case study,  \citeauthor{1967} \cite{1967} & 2019 & SelfmadeTool & DescrCorr & Java &  &  &  &  &  &  \\ 
You have said too much : Java-like verbosity anti-patterns in python codebases,  \citeauthor{n311} \cite{n311} & 2021 & ProgramQuality & DescrCorr & Python &  &  &  &  &  &  \\

\end{longtable}

\end{document}